\documentstyle [12pt,epsf] {article}

\makeatletter\@addtoreset{equation}{section}\makeatother

\def\ddk{[d^3{\rm k}]}
\def\kk{{\rm k}}

\begin{document}
\begin{titlepage}

\begin{center}
{NONEQUILIBRIUM QUANTUM DYNAMICS\\ 
OF DISORIENTED CHIRAL CONDENSATES}\\
\ \\\ \\
{\large
F. Cooper,$^1$ ,Y. Kluger,$^{1,2}$\\
E. Mottola,$^1$ and J.P. Paz$^{1,3}$\\
\ \\}
$^1$Theoretical Division and\\
$^2$ Center for Nonlinear Studies\\
Los Alamos National Laboratory\\
Los Alamos, New Mexico 87545 USA\\
\ \\
$^3$Department of Physics\\
Ciudad Universitaria\\
1428 Buenos Aires, Argentina
\ \\ \ \\

{\em ABSTRACT}
\end{center}
\quotation\noindent

The nonequilibrium dynamics of the
chiral phase transition expected during the expansion of the quark-qluon
plasma  produced in a high energy hadron or heavy ion collision is studied in
the  $O(4)$ linear sigma model to leading order in a large $N$ expansion.
  Starting from
an approximate equilibrium configuration at an initial proper time $\tau$ in
the
disordered phase we study the transition to the ordered broken symmetry
phase as the system expands and cools. We give results for the proper time
evolution of the effective pion mass, the order parameter $<\sigma>$
 as well as
for the pion two point correlation function expressed in terms of a time
dependent phase space number density and pair correlation density. We determine
the phase space of initial conditions that lead to instabilities (exponentially
growing long wave length modes) as the system
evolves in time. These instabilities are what eventually lead to disoriented
chiral condensates. In our simulations,we found that instabilities that are
formed during the initial phases of the
expansion  exist for proper times that are at most $3\,fm/c$ and lead to
condensate regions  that do not contain
large numbers of particles. The damping of instabilities is a consequence of
strong coupling.

\endquotation
\end{titlepage}

\section{Introduction}
In recent years there have been numerous investigations concerning the
possibility of forming large correlated regions of misaligned vacuum
during highly energetic collisions \cite{Anselm,BjorkenIJMP,RajaWil}.
Those regions, in which the quark
condensate $<\bar q_i~ q_j>$ is nonzero but points along the wrong direction
in isospin space
have been named Disoriented Chiral Condensates (DCC's).
If large DCC's are indeed formed, they would produce spectacular events in
which one could observe strong correlations between the emitted pions. In fact,
the  original idea of DCC's was invented \cite{Anselm} to explain rare
events observed in
cosmic ray experiments \cite{Cosmicrays}:
the Centauro events, where there is a deficit of
neutral pions. In this approach the explanation of the deficit would be
the result of the decay of domains in which the condensate  has a vanishing
component in the $\pi^0$ direction.  Anti--Centauro  events
(whose experimental status is still uncertain),  where the emission would be
predominantly neutral would be explained as  coming from regions in which the
condensate points along the $\pi^0$ direction.

If these events are a result of the creation of DCC's, this  would be
direct evidence for the existence of a chiral phase transition in the
plasma formed following an ultrarelativistic collision and would allow
us to explore the physics of  the chiral phase transition. The
possibility of producing DCC's in high energy collisions has originated
several
experimental proposals \cite{BjorkenIJMP,BjorkenKT} and a number  of
theoretical papers
\cite{Bedaque} -
\cite{BoyanDCC}.  It has been recognized that the possibility of forming  large
regions of DCC relies on the existence of a substantially large regime  in
which
the hot plasma formed after the collision evolves {\it out of equilibrium}
\cite{RajaWil}. In fact, if thermal equilibrium is approximately preserved by
the dynamics, the typical correlation length would be determined by the  pion
mass and therefore would be too small to matter. For this reason, there have
been
a number of authors making different attempts to analyze the nonequilibrium
aspects of the dynamics of the chiral phase transition.  These attempts vary in
form and content: some authors  performed numerical simulations
on classical models \cite{RajaWil,GavinPis,Huang}, others
used phenomenological terms -- inspired in classical kinetic
theory -- to model the interaction
between the condensate and the quasiparticles \cite{GavinMu} while some
attempted to incorporate quantum and thermal fluctuations
\cite{Bedaque,BoyanDCC} into the theoretical framework.

There are clearly two important questions that any theoretical
model should answer. The first one is to determine whether
during the evolution that follows the collision there
are instabilities affecting the fluctuations. If this happens,then there is a
chance for the correlations to grow. If this occurs, structure may tend
to form through a process like spinodal decomposition.
In our simulations of the chiral transition in the sigma model we find that
several types of initial fluctuations involving the time derivative of
the $\sigma$ field lead to instabilities, so that we indeed find that the
first ingredient is present in our simulations.

 The second question is  if, assuming the instability exists, the correlated
domains can grow large enough so that many pions can be emitted from each
domain
making the detection  of DCC's possible.  The time scale for the
instabilities to exist is strongly influenced by the strength of the
interactions.  In fact,
some  of the intuition one may have developed by analyzing similar problems in
other contexts, such as the cosmological one, where the coupling is extremely
weak, may not apply at all. Using the linear sigma model as the effective
field theory valid  near the chiral phase transition temperature, we find that
we
need to be in a regime of strong coupling  ($\lambda_r  \approx 10$) in order
to  fit low energy scattering data.  In the strong coupling regime
instabilities
exist only for a short period of proper time as we shall see below. This makes
it difficult to obtain large sizes for the correlated domains. On  the other
hand in a weakly coupled system the domains can grow for long periods of time
and a totally different picture can emerge.

In our approach we do not put in instabilities by hand but look at typical
fluctuations of initial conditions starting  in a region of stability. Previous
researchers have assumed the existence of  an initial instability and focused
their
calculations
on studying the growth of correlations.  The typical and quite drastic way in
which the instability had been previously introduced in the problem was by
using
the so called quench approximation. In a quench, the  system (the hot plasma
created after the collision) is subject to a sudden  external action (the
expansion into the vacuum) that has the following  net effect: it does not
change the state of the system, but instantaneously  modifies the effective
potential turning it ``upside down''. This is  clearly an idealization that may
be entirely inappropriate. In this paper  we will present an approach that does
not require this approximation and  that enables us to study (in a simplified
model) the existence of  instabilities in a rather straightforward way.

Thus, our aim is to
study the evolution of DCC's without imposing the quench approximation
or any other ad hoc way of modelling the appearance of an instability. That
is we start in a situation where the particle masses are positive and one
is in the symmetric phase. We start  at an initial value of the proper time
$\tau_0$ in the symmetric phase by choosing
a thermal distribution of particles above the critical temperature. This
initial condition is not necessary, but is one way of ensuring that
the initial system is in the disordered phase. We imagine that the system
cools via the expansion of the system into the vacuum. By cooling we
mean a reduction in the energy density with proper time. If we
were in equilibrium, the temperature would automatically decrease with
the proper time. The simplified picture we have of cooling is the  boost
invariant picture popularized by Bjorken \cite{BjorkenPRD}. This picture is
consistent with
various hydrodynamic approaches to hadronic as well as heavy ion collisions.
We kinematically constrain expectation values to only depend on the fluid
proper time $\tau = \sqrt{t^2-x^2}$, where x is the distance along the
collision axis in the
center of mass frame. The natural expansion and subsequent cooling of the
plasma into the ordered vacuum is studied numerically by solving
the update equations for the quantum modes as well as for the proper
time evolving expectation values. In this way, we are able to study the
evolution
of the plasma in a self consistent way without imposing any instability by
hand.
We analyze various reasonable initial conditions on the fields and determine
whether they lead to  instabilities. That is for various initial conditions
we determine the proper time evolution of the effective pion mass. When the
effective pion mass becomes negative  instabilities ensue. We also determine
the time evolving order parameter $<\sigma>$ and the adiabatic phase space
number
density and pair density which determine the spatial correlation function for
the pion field. These number densities are related to physical measurables
(such
as the rapidity distribution of final particles) at later times. The initial
conditions which lead to the largest instabilities have initial velocities in
either the $\sigma$ or $\pi$ directions. We compare the nonequilibrium results
for the number density with those that  would have resulted from an expansion
with local thermal equilibrium in the comoving frame. When instabilities
arise the distributions tend to narrow in momentum space, especially in the
transverse direction.

In the investigations done so far, simple phenomenological
models have been used hoping that they
describe the fundamental physics involved in the dynamics of the
phase transition. We will employ
the linear sigma model, the most popular one in this context,
which seems to have the essential
attributes of being simple but realistic enough: it appropriately
describes the low energy phenomenology of pions and has also the correct
chiral symmetry properties. The initial  conditions we will impose  are
motivated by matching it to the situation one expects to attain in a
highly energetic collision. We will assume that the quantum state of the system
is a thermal density matrix at an initial instant of proper time $\tau_0$. We
will choose the initial temperature $T$ to be slightly above the critical
temperature for being in the disordered phase. As shown in the appendix
the critical temperature is given by: $ T_c^2  = 3 f_{\pi}^2$.  We
will choose the parameters of the model to give reasonable values for three
experimentally determined quantities : the mass of the pion $m_{\pi}$, the pion
decay
constant $f_\pi$ and the  s wave $\pi-\pi$ phase shifts above threshold. These
three measurements completely determine the parameters of the model.  One
important constraint on this model is the triviality of the model as the cutoff
is removed. The theory only makes sense at cutoffs below the Landau pole which
occurs at a value of the cutoff $\Lambda$ when the bare coupling constant first
becomes negative for positive renormalized coupling constant.  This
limits the size of the renormalized coupling constant.The maximum renormalized
coupling constant as a function of $ \Lambda$  obeys for large $\Lambda$

\begin{eqnarray}
   \lambda_r ^{max} =  {2\pi^2 \over {\rm ln} ({2 \Lambda \over m})}
\end{eqnarray}
Since the mass difference between  the $\sigma$ and $\pi$ is directly
proportional to $\lambda_r$,
this leads to an upper bound for the
the mass of the $\sigma$ resonance as a function of $\Lambda$. Therefore,
unlike at tree level, the mass of the  $\sigma$ in the fully quantized theory
is
constrained in this model. $\Lambda$ is also constrained from the physical
considerations that we want the mass of the  $\sigma$ to be less than the
cutoff.
However the mass of the  $\sigma$ resonance increases as we decrease the cutoff
since then the renormalized coupling increases. This pins down the cutoff to
lie
between $700 MeV$ and $1 GeV$. The rest of the paper is organized as follows.
In section 2 we describe the linear sigma model in the leading order in the
$1/N$ expansion. In section 3 we discuss the Baked Alaska Scenario and we
derive renormalized update equations for the proper time evolution of
the field theory. In section 4 we describe the initial conditions we
use to study the development of disoriented chiral condensates. In section
5 we discuss the results of extensive numerical simulations.  In the appendix
we rederive for completeness the properties of the linear sigma model in the
large N approximation.

\section{Model and approximations}
We will use the $O(4)$ linear sigma model to describe the evolution
of the pions. We are well aware of the limitations of this
approach, that provides a reasonable phenomenological model only for a
limited range
of energies (typically smaller than $1\ GeV$). In spite of its shortcomings,
this model captures some of the essential physics
involved in the dynamics of the phase transition that may produce
disoriented chiral condensates. In particular the chiral phase transition
takes place at a reasonable temperature ($T_c ={\sqrt 3} f_\pi$ ), and
the low energy $\pi$ - $\pi$ scattering amplitudes are reasonable in this
model.
The mesons are organized in an $O(4)$ vector
$\Phi=(\sigma,\vec\pi)$ and the action is (in natural units $\hbar=c=1$)
\begin{eqnarray}
S=\int d^4x \{{1\over 2} \partial\Phi \cdot \partial\Phi - {1\over 4}
\lambda (\Phi \cdot \Phi - v^2)^2 + H\sigma\}.\label{action}
\end{eqnarray}
where we have use the Bjorken and Drell metric:
$(1,-1,-1,-1)$.
We will describe the evolution of the mean value
$\bar\Phi\equiv<\Phi>$ and the two point correlation functions including the
effects of quantum and thermal fluctuations. For this purpose, we
will treat (\ref{action}) as a quantum theory defining initial values (i.e.,
the
quantum state of the system at a given proper time) and will follow the real
proper time  evolution of $\bar\Phi$ and the correlations.
The complexity of the problem forces us to adopt
some approximations. Perturbation theory is useless for our purpose
\cite{BoyaPRD,BoyaPRE} and a scheme which is non--perturbative in $\lambda$
must be adopted. In this context, the most popular approaches which allow for a
real time analysis are the
the Hartree (or Gaussian) ansatz \cite{Hartree} and the large $N$
expansion of
the $O(N)$ sigma model \cite{1/n}. We will adopt the later since
it presents several advantages. On the
one hand, the expansion is systematic and allows us to study higher
order corrections (work is in progress in this direction and results will be
presented elsewhere \cite{usnext}).
On the other hand, when using the Hartree ansatz in the context of the
study of DCC's one is forced to take also the large $N$ limit (see Ref.
\cite{BoyanDCC}).
This is due to the well known fact that the Gaussian approximation violates the
Goldstone theorem giving an unphysical (and not necessarily small) mass
to the pions in the $H=0$ limit. Of course, the approximation we adopt here
is not expected to capture all the features of the phase transition (as is
well known, mean field theory fails to predict the correct critical exponents
but allows us to explore the strong coupling regime).
Because of the
triviality of the $O(N)$ sigma model as one takes away the cutoff, an aspect of
the exact theory that is preserved in the large $N$ approximation, we have to
seriously take into account the cutoff and its ramifications. One of these
ramifications is that the renormalized coupling constant
has a modest upper bound of the order of ten  at a cutoff of one GeV. This
gives an upper bound to the mass of the $\sigma$ resonance whose value depends
on our choice of cutoff.

The large $N$ effective equations can be obtained in a variety of ways, which
are extensively discussed
in the literature \cite{1/n}. A very convenient method is to
use an effective action, which is a functional of the mean values of the
original fields $\Phi$ and of an auxiliary
constrained field $\chi$ \cite{higher}. We start with a
classical action $\tilde S[\Phi,\chi]$ constructed from (\ref{action})
by replacing $\chi=\lambda(\Phi^2/2N - v^2)$. As this action is now
quadratic in $\Phi$ we can perform the functional integral over those
fields and are left with a functional integration over $\chi$ which, to
leading order in $1/N$ can be calculated by the stationary phase method.
In the appendix we review the details of this calculation and calculate all
the propagators and the $\pi-\pi$ scattering amplitude in the leading order in
the large N  expansion. Higher
order corrections can be systematically computed in this way \cite{higher} and
an expansion of the effective action $\Gamma[\bar\Phi,\bar\chi]$ in powers
of $1/N$ can be obtained \cite{1/n}. We will consider here only the leading
order terms which give
the following  equations (for notational convenience we drop the overbars
and denote the expectation values  $\Phi$ and $\chi$):
\begin{eqnarray}
\bigl(\Box_x +
\chi(x)\bigr) \Phi_i(x) &=& H\delta_{i1}\label{phieq}\\
\chi(x) &=& \lambda\bigl(-v^2+\Phi^2(x)+N G_0(x,x)\bigr)\label{chieq}\\
G_0^{-1}(x,y)&=& i\bigl(\Box_x + \chi(x)\bigr)\delta(x-y).\label{ginverse}
\end{eqnarray}

The structure of these equations is indeed very simple. The field $\chi$
plays the role of the effective mass for the mean values $\Phi_i$ and satisfies
  the ``gap equation '' (\ref{chieq}). The function $G_0(x,x)$ that appears in
(\ref{chieq}) is
the coincidence limit of the propagator $G_0(x,y)$ that inverts the operator
$G_0^{-1}$  defined in (\ref{ginverse}). We can use an auxiliary quantum field
$\phi(x)$ where  $<\phi(x)>=0$ and
\begin{eqnarray}
\bigl(\Box_x +
\chi(x)\bigr) \phi(x)=0
\end{eqnarray}
to determine $G_0(x,y)$.We construct the propagator as
$$G_0(x,y)=< T~\phi(x)\phi(y)>$$.
$ \delta_{ij} G_0(x,y)$ is the  the pion propagator when
$<\pi_i(x)>=0$.
The initial value problem
associated with equations (\ref{phieq})--(\ref{ginverse}) will be solved in the
next section.
Here, we would like to address the issue of how to use this model to make
contact with the phenomenology we want to describe. Thus, we must fix the
values
of the parameters appearing in the above equations so that they describe low
energy pion  physics. The measurable quantities we want to reproduce are the
pion mass  $m_\pi=135\ MeV$, the pion decay  constant $f_\pi=92.5\ MeV$  as
well
as the s wave, $I=0$ phase shifts in the energy range $300-420 MeV$.
To fit these physical quantities we  analyze  our equations in the ``true
vacuum'' state (i.e., in equilibrium at zero temperature). In such state the
derivatives of the expectation values vanish and we have  $\Phi=(\sigma_v,\vec
o)$, $\chi=\chi_v$ where $\sigma_v$ and $\chi_v$ are some  constants whose
values
will be determined below.   The physical masses can be related to the
parameters
of the theory by computing the inverse propagators of the pion and sigma
fields. The s-wave phase shifts is determined from the $\pi-\pi$ scattering
amplitude obtained in this approximation which is given by the
exchange of the composite field $\chi$ propagator in all three channels. A
complete review of
the vacuum and finite temperature properties of this model in leading
order in large N is found in the
appendix and we summarize the results here. The vacuum expectation value of
$\sigma$ is determined by $f_\pi$
\begin{eqnarray}
\sigma_v=f_\pi.\label{fpisigma}
\end{eqnarray}
On the other hand, in the vacuum we have $\vec\pi=0$ and the pion inverse
propagator is
$G^{-1}_{ij~\pi,\pi}(x,y)=G_0^{-1}(x,y) \delta_{ij}$. Therefore,
the vacuum expectation value of $\chi$ is
\begin{eqnarray}
\chi_v=m_\pi^2 \equiv m^2.\label{mpichi}
\end{eqnarray}
The sigma mass can be  approximately determined in terms of the
inverse  sigma propagator as the zero in the real part of the inverse
propagator
(a more precise determination which gives a slightly different result is from
the peak in the $I=0$, $l=0$ scattering amplitude).
This leads to the equation:

\begin{eqnarray}
m_{\sigma}^2=m^2 +\sigma^2 Re[\hat{G}_{0 ~\chi
\chi}(m_{\sigma}^2)],\label{msigmadef}
\end{eqnarray}
where  $\hat{G}_{0 ~\chi \chi}$ is the composite field propagator in the
absence
of symmetry breaking, i.e.
\begin{eqnarray}
{\hat{G}^{-1}}_{0 \chi \chi}(p^2) = { 1 \over  2 \lambda} + {N \over 2}
\Pi(p^2),
\label{ginchi}
\end{eqnarray}
and the  polarization $\Pi = i G_{0}^2$ is given by
\begin{eqnarray}
\Pi(p^2) = -i \int ~[d^4q ]~ (\chi-q^2)^{-1} (\chi-(p+q)^2)^{-1},\label{pol}
\end{eqnarray}
and is  explicitly given in the appendix. We use the notation that
$[d^n q] = d^n q/(2 \pi)^n$.
In the absence of symmetry breaking the composite field
propagator is the geometric sum of the bubble chains. In the presence
of symmetry breakdown one is also summing the graphs where the bubble
is replaced by a single
propagator and two tadpoles (this includes the sigma exchange graphs). One can
determine the bare mass of the pion in terms
of the physical pion mass using the  gap equation, since $\chi_v = m^2$

\begin{eqnarray}
m^2 = - \lambda v^2 + \lambda f_{\pi}^2 + \lambda N G_0 (x,x)
\end{eqnarray}
where
\begin{eqnarray}
G_0(x,x) =\int_0^{\Lambda} \ddk {1\over 2\sqrt{k^2+m^2}}
\end{eqnarray}

This relationship tells us that the bare mass goes to negative infinity
quadratically with the cutoff, as is true for the exact lattice theory.
(That is the reason why in strong coupling the theory is related to the
Ising Model). This relationship also allows us to eliminate the bare mass
from the theory as well as the quadratic dependence on the cutoff.
The logarithmic dependence on the cutoff are removed by coupling constant
renormalization.
We obtain a value of the bare coupling $\lambda$ at a fixed cutoff $\Lambda$
(or equivalently the renormalized coupling $\lambda_r$) by comparing our
large-N result with the Pade fit of Basdevant and Lee \cite{BasLee} to the
$I=0$
s-wave scattering amplitude.  This is discussed in detail in the appendix.
Once
$\lambda$ is determined both $v$ and $m_{\sigma}$ are fixed.
Therefore all the bare parameters of this theory are fitted in terms of
the three experiments which determine $m$, $f_{\pi}$, and $\lambda$.
 The external field $H$ is determined from the equilibrium time independent
solution to field equation for the $\sigma$ field:  $ \chi \sigma= m_{\pi}^2
f_{\pi} =H$.
  Our numerical simulations  automatically have a  cutoff since we
are performing numerical simulations in a box. Thus one can use either the
bare or renormalized parameters to describe the problem.  The renormalized
parameters, however  are determined from physical measurements and are more
fundamental. However since the $\sigma$ model is not an asymptotically free
field theory, it is a trivial theory when we take the cutoff away. Thus one is
not allowed to take the continuum limit and must  consider the theory as an
effective field  theory which is not valid for energies in excess  of a few
$GeV$'s. Our model is  a theory which makes physical sense (such as having a
well defined ground state) only with a physical cutoff $\Lambda_p$ (see
\cite{cutoff} \cite{Schnitzer} for discussions on  $\Phi^4$ as a cutoff
theory).
   On physical grounds we want $ 2 m_{\pi} < m_{\sigma} < \Lambda$. This
relation is quite a constraint on $\Lambda$ since when we lower $\Lambda$
we increase the value of the mass of the sigma.  We
do not  expect the theory to be valid for energies above 1 GeV since in
that regime the correct dynamics is described by QCD.   In
fact if we raise $\Lambda$ above 1 GeV, the maximum value of the renormalized
coupling goes down and the $\sigma$ mass becomes unacceptably low.
Reasonable values for the mass of the $\sigma$ constrain $\Lambda$ to be
in the range $ .7 GeV  < \Lambda \leq 1 GeV$.  In that range the best
value of the renormalized coupling $\lambda_r$ is between 7-10.
For this range of values for $\lambda_r $ this model agrees qualitatively with
low energy scattering data. With this choice of parameters,
 it is difficult to obtain values of the $\sigma$ mass higher
than $450 MeV$.
Based on the perturbative calculations of Basdevant and
Lee \cite{BasLee} which were then subjected to a Pade improvement, and the
connection between $1/N$ expansions and resummations  we
expect that at next order in $1/N$ this upper bound on the $\sigma$ mass will
be
raised slightly.

As will be clear from our discussion below, when solving the
equations we can use a different value of the cutoff
provided we scale the bare couplings appropriately (so that we keep the
physical quantities unchanged). As the theory we are
using is renormalizable, this scaling is well defined and
known. However, the cutoff can not
be taken too large since the theory only makes sense for positive
values of the bare coupling. This is the Landau pole problem and is
related to the triviality of the theory. \cite{Baker} \cite{bendercooper}.
As is discussed in detail in the appendix, the bare and renormalized couplings
are related by the equation:
\begin{eqnarray}
\lambda = { \lambda_r \over 1- N \lambda_r \Pi (0)} \label{renorm1}.
\end{eqnarray}
We see that $\lambda$ has a pole when $ 1= N \lambda_r \Pi (0)$.
At large values of the  cutoff this Landau pole occurs
  when
\begin{eqnarray}
\log(2  \Lambda/m_\pi) \approx
{8\pi^2 \over  N \lambda_r}.\label{cutoff}
\end{eqnarray}
For fixed $\lambda_r$  we notice that for values of the cutoff larger than that
given by (\ref{cutoff}), the
bare coupling becomes negative which makes the theory undefinable as a
Euclidean
lattice field theory.
This relationship also  means that at a fixed cutoff, there is a maximum
$\lambda_r$ that we can consider. That is the inverse relationship is
\begin{eqnarray}
\lambda_r = { \lambda \over 1+ N \lambda \Pi (0)} \label{renorm2}.
\end{eqnarray}
The maximum of $\lambda_r$ is reached at infinite $\lambda$ and one obtains:
\begin{eqnarray}
\lambda_r^{max}  = { 1 \over   N \Pi (0)} \label{max}.
\end{eqnarray}
 For a cutoff of
$1 GeV$ the maximum $\lambda_r$ is 13.

As a final comment we would like to point out that the effective
action method we employed can also be used to compute
higher order corrections in $1/N$. In that case it is necessary to use a
formalism
that enable us to derive real and causal equations for the expectation values
(otherwise
one cannot even pose the initial value problem). This formalism
is known as the ``closed time path'' method developed by Schwinger,
Mahanthappa,
and Keldysh \cite{CTP}. The analysis  of the $1/N$ corrections obtained using
this approach (that carry the effects of  the collisions that may produce
relaxation towards equilibrium) will be analyzed elsewhere  \cite{usnext} (see
also Ref. \cite{Bedaque,BoyaPRD,1/n} for recent applications of this formalism
in the
DCC and other related contexts).

\section{Cooling by expansion and Baked Alaska scenario}

\subsection {The basic idea}

\bigskip
To analyze the possibility of forming DCC's we should take into account
the specific features characterizing the situation after a highly
energetic heavy ion collision. Experimentally, a flat plateau in the
distribution of
produced particles per unit rapidity is observed in the central rapidity region
(these results are  obtained in $p$--$\bar p$ and other collisions).
This suggest the existence of an
approximate Lorentz boost invariance. Thus, the simplest picture of a
collision, due to Landau \cite{Landau},
is one in which the excited nuclei are highly contracted
pancakes receding  away from the collision point
at approximately the speed of light.
The boost invariance implies that the evolution of the
``hot plasma'' that is left in between the nuclei looks the same
when viewed from different inertial frames. Of course, this is an
approximate picture that is not valid for large values of the spatial rapidity
and for transverse distances of the order of the nucleus size.
The
existence of this approximate symmetry can be used to make a very simple
hydrodynamical model \cite{BjorkenPRD}
that, in some cases, may describe the evolution of
the plasma. It is worth reviewing very briefly these ideas.
First one should recognize that the natural coordinates
to make a  boost invariant model are the
proper time $\tau$ and the spatial rapidity
$\eta$ defined as
\begin{eqnarray}
\tau\equiv(t^2-x^2)^{1/2}, \qquad \eta\equiv{1\over 2}
\log({{t-x}\over{t+x}}).
\label{taueta}
\end{eqnarray}
The observed symmetry will be respected by the model if one imposes initial
values
on a $\tau=$constant hypersurface (and not at constant laboratory time $t$).

If we had local thermodynamic equilibrium during the expansion so that
the relaxation rate is faster than the expansion rate and assumed
that we could approximate the field theory dynamics with a hydrodynamic
flow, one would find by
solving the hydrodynamical equations with this type of initial
conditions (homogeneity along the constant $\tau$ surface)
 that the energy density drops as $\tau^{-\alpha}$. Here
$\alpha=1+c_0^2$ where $c_0$ is the speed of sound, which depends on
the equation of state of the fluid $p=c_0^2 \epsilon$. In the ultrarelativistic
case $c_0^2 = 1/3$ and the temperature falls as $\tau^{-1/3}$.

According to this simple model of a collision, the plasma evolves
in a highly inhomogeneous way when viewed from the laboratory frame.
In fact, analyzing a constant $t$ surface we realize that the
field configurations strongly depend on the spatial coordinate $x$.
Near the light cone $|x|=t$ the system is ``hot'' (corresponding
to small values of the proper time $\tau$). On the contrary,
for small values of $x$ (that correspond to larger values of $\tau$)
the system is ``colder''. This type of configuration, hot in the
outside and cold in the inside, is schematically known as Baked Alaska
(or a Swiss Bombon, according to other culinary traditions).

In this paper we want to study the formation of DCC's using some of
the ideas presented above. We {\it will not} assume a
quasi equilibrium situation or use a phenomenological hydrodynamical
(or kinetic) model to describe the evolution of our system. On the
contrary, we will
study the nonequilibrium evolution in its full glory and solve
equations (\ref{phieq})--(\ref{ginverse}), which include thermal and quantum
effects.
Using the coordinates (\ref{taueta})
and fixing boost invariant initial conditions at an initial proper time
$\tau_0$ (whose numerical value we discuss below) we introduce the expansion
and ``cooling'' of
the plasma in a natural way. Therefore, we do not need to introduce any ad hoc
cooling mechanism by hand. The cooling, if any, will
appear as a result of the evolution, which is fully out of equilibrium.
In this way, we can really test the
validity of the ``quenching'' approximation that has been almost
universally used when analyzing the evolution of DCC's.

\subsection{The equations}
\bigskip

Our treatment is very similar to the one required to study quantum field theory
in a curved spacetime \cite{BirrellDavies}. Thus, in the coordinates
(\ref{taueta})
Minkowsky's arc element is
\begin{eqnarray}
ds^2=d\tau^2-\tau^2d\eta^2-dx_\perp^2,\label{metric}
\end{eqnarray}
which has the same form of a Kasner Universe (an anisotropically expanding
Universe,
which in this case is nevertheless flat, see \cite{BirrellDavies}).
To study the evolution of the mean fields and correlations in
this coordinates our first task is to rewrite equations
(\ref{phieq})--(\ref{ginverse}) using
the new variables. Assuming that the mean values $\Phi$ and $\chi$ are
functions
of $\tau$ only (homogeneity in the constant $\tau$ hypersurface) we have
\begin{eqnarray}
\tau^{-1}\partial_\tau\ \tau\partial_\tau\ \Phi_i(\tau) +\ \chi(\tau)\
\Phi_i(\tau) =\
H\delta_{i1}\label{phieq2}
\end{eqnarray}
\begin{eqnarray}
\chi(\tau) = \lambda\bigl(-v^2+\Phi_i^2(\tau)+
N <\phi^2(x,\tau)>\bigr)\label{chieq2}
\end{eqnarray}
where the quantum field $\phi(x,\tau)$ satisfies the Klein Gordon equation
\begin{eqnarray}
\Bigl(\tau^{-1}\partial_\tau\ \tau\partial_\tau\ - \tau^{-2}\partial^2_\eta
-\partial^2_\perp + \chi(x)\Bigr)
\phi(x,\tau)=0.\label{fieldeq2}
\end{eqnarray}
The quantum field $\phi(x,\tau)$ defined here is an auxiliary field which
allows us to calculate the Wightman function $G_0 (x,y)$ by taking the
expectation value:
\begin{eqnarray}
    G_0 (x,y;\tau) \equiv < \phi(x,\tau) ~\phi(y,\tau)>.
\end{eqnarray}
When the expectation value of the pion field is zero, then this field
corresponds to one component of the pion field.

As usual, we expand this field in an orthonormal basis
\begin{eqnarray}
\phi(\eta,x_\perp,\tau)\equiv{1\over{\tau^{1/2}}} \int \ddk\bigl(\exp(i\kk
\cdot
{\bf x}) f_\kk(\tau)\ a_\kk\ + h.c.\bigr)\label{fieldexp}
\end{eqnarray}
where $\kk \cdot
{\bf x}\equiv k_\eta \eta+\vec k_\perp \vec x_\perp$, $\ddk\equiv dk_\eta
d^2k_\perp/
(2\pi)^3$
and the mode functions $f_\kk(\tau)$ evolve according to (a dot here
denotes the derivative with respect to the proper time $\tau$):
\begin{eqnarray}
\ddot f_\kk +
\bigl({k_\eta^2\over{\tau^2}}+\vec k_\perp^2 + \chi(\tau) +
{1\over{4\tau^2}}\bigr)
f_\kk=0.\label{modeq}
\end{eqnarray}
For the creation and annihilation operators to obey the usual canonical
commutation relations $ [a_k,a^{\dag}_k] = 1$, then the mode functions
must satisfy the Wronskian condition:
$$
  f \dot{f}^{\ast} - f^{\ast} \dot{f} = i
$$
The expectation value $<\phi^2(x,\tau)>$ can be expressed in terms of the mode
functions $f_\kk$ and of the distribution functions
\begin{eqnarray}
n_\kk\equiv <a^\dagger_\kk a_\kk>, ~~
g_\kk\equiv <a_\kk a_\kk>,
\label{dist}
\end{eqnarray}
 which entirely
characterize the initial state of the quantum field. For simplicity, we will
assume
that the initial state is described by a density matrix which is diagonal in
the
number basis (like a thermal state). In such case,
the only nonvanishing distribution is $n_\kk$. Thus, replacing the above
expressions
in (\ref{chieq2}) we have:
\begin{eqnarray}
\chi(\tau) = \lambda\Bigl(-v^2+\Phi_i^2(\tau)+{1\over \tau}N \int \ddk
|f_\kk(\tau)|^2\  (1+2\ n_\kk) \Bigr).\label{chieq3}
\end{eqnarray}

We assume the initial density matrix is one of local thermal equilibrium
in the comoving frame.  In the comoving frame, the expectation value
of the energy momentum tensor is diagonal and of the form
\begin{eqnarray}
{\rm diag} (\epsilon,p,p,p).
\end{eqnarray}
 We then have at $\tau=\tau_0$ (the surface
of constant energy density and temperature $T_0$) that:
\begin{eqnarray}
n_k = {1 \over e^{\beta_0  E^0 _k} -1} \label{initn},
\end{eqnarray}
where $ \beta_0 = 1/T_0$, $E^0_k= \sqrt{k^2(\tau_0)+\chi(\tau_0)}$ and $k^2 =
k_\eta^2/\tau^2 + k_{\perp}^2$.

We can also use a time dependent set of creation and annihilation operators
to describe the quantum field $\phi$.  If we use the first order
adiabatic mode function:
\begin{eqnarray}
f^0_k = {e^{-iy_k(\tau)} \over \sqrt{2 \omega_k}}; ~~ dy_k/d \tau =
\omega_k,\label{admode}
\end{eqnarray}
where $\omega_\kk(\tau)\equiv({k_\eta^2+1/4 \over \tau^2}+
\vec k_\perp^2 + \chi(\tau))^{1/2}$
Then if we expand the field in terms of these mode functions we h
\begin{eqnarray}
\phi(\eta,x_\perp,\tau)\equiv{1\over{\tau^{1/2}}} \int \ddk\bigl(\exp(i\kk
{\bf{x}})
f^0_\kk(\tau)\ a_\kk(\tau)\ + h.c.\bigr)\label{adfieldexp}.
\end{eqnarray}
We can now define the first order adiabatic interpolating number and
pair distribution functions via:
\begin{eqnarray}
n_\kk(\tau) \equiv <a^\dagger_\kk(\tau)  a_\kk (\tau)>, ~~
g_\kk (\tau)\equiv <a_\kk(\tau) a_\kk(\tau)>\label{disttau}
\end{eqnarray}

The time dependent creation and annihilation operators satisfy
\begin{eqnarray}
\dot{a}f^0 + \dot{a}^{\dag} f^{*0} = 0
\end{eqnarray}
in order that the usual canonical commutation relations hold.
Then we can rewrite the Green's function $G^0$ in the two bases:
\begin{eqnarray}
G^0(x,y;\tau)  = {1\over \tau} \int \ddk e^{ik(x-y)}
{1 \over 2 \omega_k(\tau)} \  (1+2\ n_\kk (\tau) + 2 Re [g_k(\tau)
e^{-i2y_k(\tau)}]) \nonumber \\
\end{eqnarray}
or
\begin{eqnarray}
G^0(x,y;\tau)  = {1\over \tau} \int \ddk e^{ik(x-y)} \  [1+2\ n_\kk (0)] ~~
\vert
f_k(\tau)\vert^2.
 \end{eqnarray}

The $\tau$ dependent  variables $n$ and $g$ now have the physical
interpretation
as the interpolating phase space number and pair density for the case when
$< \pi_i> =0$. These operators agree with the time independent number and pair
densities defined by (\ref{dist})
at $\tau_0$, and in the out regime become the physically measurable number
and pair densities. For $<\pi_i> \neq 0$ the actual pion
two point function is  more complicated than $G^0$ and this is only one piece
of that Green's function. Of course in the vacuum Isospin conservation
requires $<\pi_i> = 0$, so that once we are approaching the final true
vacuum state during the cooling process, one can use these interpolating
number and pair operators to describe  the physics of the problem.
Also note that during the instability phase, $\omega_k^2$ can become negative
and for those modes the adiabatic basis does not exist because the
Wronskian condition for the mode functions is no longer satisfied.

The two sets of creation and annihilation operators are connected by
a Bogoliubov transformation:
\begin{eqnarray}
a_k(\tau) = \alpha(k,\tau) a_k + \beta(k,\tau) a^{\dag}_{-k}.
\end{eqnarray}
$\alpha$ and $\beta$ can be determined from the exact time evolving mode
functions via:
\begin{eqnarray}
\alpha(k,\tau) = i (f_k^{0\ast} { \partial f_k \over \partial \tau} -
{\partial f_k^{0 \ast}  \over \partial \tau}f_k)
\end{eqnarray}
\begin{eqnarray}
\beta(k,\tau) = i (f_k^{0} { \partial f_k \over \partial \tau} -
{\partial f_k^{0 }  \over \partial \tau}f_k)
\end{eqnarray}

In terms of the initial distribution of particles $n_k(\tau_0)\equiv n_k$ and
the Bogoliubov coefficients $\alpha$ and $\beta$
we have:
\begin{eqnarray}
n_k(\tau) = n_k + |\beta(k,\tau)|^2 ( 1+2n_k)
\end{eqnarray}
\begin{eqnarray}
g_k (\tau) = \alpha(k,\tau) \beta(k,\tau) ( 1+2n_k)
\end{eqnarray}
  \subsection{Initial conditions and renormalization}
\bigskip We desire to solve equations  (\ref{phieq2}),(\ref{modeq}) and
(\ref{chieq3})
as an initial value problem. To do this we need to give Cauchy data (the
function  and its derivatives) for the mean values $\Phi_i(\tau)$ and for the
mode functions $f_\kk(\tau)$. This, together with the distribution function
$n_\kk$ and $g_\kk$ at $\tau_0$ fully defines the problem. Notice that the
initial value of $\chi(\tau)$ is determined self-consistently by solving the
gap
equation (\ref{chieq3}) at the initial time $\tau_0$.

The quantum state is fully determined by the distribution $n_\kk$ and by the
initial data $f_\kk(\tau_0)$, $\dot f_\kk(\tau_0)$. These data fix the vacuum
state upon which the Fock space is built. The definition of the vacuum state in
a  curved spacetime is certainly a tricky issue with a rather long history
\cite{BirrellDavies}. Fortunately this is not an important problem for us,  due
to the fact that the background spacetime is flat. Thus, it is possible to
choose the initial  data $f_\kk(\tau_0)$ and $\dot f_\kk(\tau_0)$ so that the
vacuum state coincides with the ordinary Minkowsky vacuum, at least for high
momentum. It can be shown that  this is accomplished by taking the ``zero order
adiabatic'' vacuum  where \begin{eqnarray}
f_\kk(\tau_0)={1\over{2\omega_\kk(\tau_0)}},\qquad \dot f_\kk(\tau_0)=(i\
\omega_\kk(\tau_0)
-{\dot\omega_\kk(\tau_0)\over{2\omega_\kk(\tau_0)}})f_\kk(\tau_0),\label{finitial}
\end{eqnarray} where $\omega_\kk(\tau)\equiv({k_\eta^2/{\tau^2}}+\vec k_\perp^2
+ \chi(\tau))^{1/2}$.  We will fix these initial conditions for the normal
modes, which are normalized according  to the Wronskian condition $f^*\dot
f-\dot f^* f=i$. The distribution $n_\kk$ will be taken as thermal,
characterized by a temperature $T$: \begin{eqnarray}
n_\kk=\bigl[\exp(\omega_\kk(\tau_0)/k_BT)-1]^{-1}.\label{thermaldist}
\end{eqnarray}

As we said before, the existence of ultraviolet divergences is not a very
delicate
issue here since we are dealing with a theory that has a natural cutoff.
However,
it is worth mentioning here how the removal of divergences can be implemented.
For
the initial conditions (\ref{finitial}), the divergences in $<\phi^2(x,\tau)>$
can be
shown to be the same ones obtained in the lowest order adiabatic approximation
, i.e.:
\begin{eqnarray}
<\phi^2(x,\tau)>_{div}={1\over \tau}\int \ddk
{1\over{2 \omega_\kk(\tau)}}.\label{phisquarediv}
\end{eqnarray}
Introducing a cutoff at the physical momenta $k_\eta/\tau$ and $k_\perp$, we
can easily
see that the above integral has both quadratic and logarithmic divergences. To
renormalize
the theory we absorb the quadratic divergences in the bare mass $\lambda v^2$
and
the logarithmic ones in the coupling constant $\lambda$. After a few simple
manipulations (that involve adding and subtracting the appropriate terms in
(\ref{chieq3}))
we can write the equation for $\chi$ as
\begin{eqnarray}
\chi(\tau)= \chi(\tau_0)&+&{\lambda_r\over
Z}\bigl(\Phi_i^2(\tau)-\Phi_i^2(\tau_0)\bigr)
+{N\lambda_r\over \tau Z}\int\ddk\Bigl\{|f_\kk(\tau)|^2(1+2\ n_\kk)-\nonumber\\
&-&{1\over{2\tilde\omega_\kk(\tau)}}\Bigr\}-{N\lambda_r\over \tau_0
Z}\int\ddk{1\over{\omega_\kk(\tau_0)}}n_\kk,\label{chieqren}
\end{eqnarray}
where $\tilde\omega_\kk(\tau)\equiv(k_\eta^2/{\tau^2}+\vec k_\perp^2 +
\chi(\tau_0))^{1/2}$ and
the renormalized coupling $\lambda_r$ and $Z$ are defined as:
\begin{eqnarray}
\lambda=\lambda_r/Z,\qquad Z=1-\lambda_r\delta\lambda,\qquad
\delta\lambda\equiv{N\over 4\tau}\int\ddk{1\over{\tilde\omega_\kk^3(\tau)}}
.\label{lambdaz}
\end{eqnarray}

The  initial value $\chi(\tau_0)$ comes from solving the gap equation
(\ref{chieq3}) at  the initial time.
As we discussed earlier it is useful to rewrite eq. (\ref{lambdaz}) in order to
discuss triviality.  Namely we have the inverse equation:
 \begin{eqnarray}
 \lambda_r ={  \lambda \over 1 + \delta \lambda  ~ \lambda}
\end{eqnarray}
  The result is that at a fixed value of $\Lambda$ there
is a maximum renormalized coupling which decreases with the logarithm of the
the cutoff. As we take the cutoff away, the renormalized
coupling constant goes to zero, signifying the triviality of the theory
as we remove the cutoff as first discussed by Baker and Kincaid  \cite{Baker}
and Bender et. al. \cite{bendercooper}

To make our presentation more clear it may be useful to
put together the set of equations we will solve. They are
\begin{eqnarray}
&&\ddot \Phi_i(\tau)+\tau^{-1}\dot \Phi_i(\tau) +\ \chi(\tau)\ \Phi_i(\tau) =\
H\delta_{i1},\label{finaleqphi}\\
&&\chi(\tau)= \chi(\tau_0)+{\lambda_r\over
Z}\bigl(\Phi_i^2(\tau)-\Phi_i^2(\tau_0)\bigr)
+{N\lambda_r\over \tau Z}\int\ddk\Bigl\{|f_\kk(\tau)|^2(1+2\ n_\kk)-\nonumber\\
&&\qquad\qquad-{1\over{2\tilde\omega_\kk(\tau)}}\Bigr\}-{N\lambda_r\over \tau_0
Z}\int\ddk{1\over{\omega_\kk(\tau_0)}}n_\kk,\label{finaleqchi} \\
&&\ddot f_\kk +
\bigl({k_\eta^2\over{\tau^2}}+\vec k_\perp^2 + \chi(\tau) +
{1\over{4\tau^2}}\bigr)
f_\kk=0.\label{finaleqmode}
\end{eqnarray}

It is worth noticing that in these equations the ``bare parameters'' $\lambda$
and $v^2$ do not appear.
Thus, the equations are entirely written in terms of renormalized
quantities $\lambda_r$ and $\chi_0$. The value of $\chi_0$
 is obtained from the gap equation (with
the same cutoff in all the integrals). Once this
is done, the value of the cutoff can be changed at will in
(\ref{finaleqphi})--(\ref{finaleqmode})
(provided we don't cross the Landau pole).

For future use, it is convenient to compute the initial value of $\dot\chi$,
which is determined by equation (\ref{finaleqchi}). It simply reads
\begin{eqnarray}
\dot\chi(\tau_0)\Bigl(1+{N\lambda_r\over{2\tau_0}}\int\ddk {n(\kk)\over
{\omega_\kk^3(\tau_0)}}\Bigr)&=&
2\lambda_r\Phi_i(\tau_0)\dot\Phi_i(\tau_0)\nonumber\\
&-&{N\lambda_r\over{\tau_0^2}}\int\ddk {n(\kk)\over{\omega_\kk(\tau_0)}}
\bigl(1-{k_\eta^2\over{\tau_0^2\omega_\kk^2}}\bigr)\label{chidot} \nonumber \\
\end{eqnarray}
{}From here we can see that $\dot\chi(\tau_0)$ has two contributions. The
first term in the r.h.s. of (\ref{chidot}) carries the contribution of the
mean values: as expected, $\dot\chi(\tau_0)$ is sensitive to
the existence of initial velocities.
The second term, which is always negative, arises from the
finite temperature part of the initial state. It clearly shows how the
expansion tends to reduce the initial value of $\chi$.

It is instructive at this point to look at the simpler equations that arise
in flat space for a spatially homogeneous expansion. In that case we have:

\begin{eqnarray}
\ddot \Phi_i(t)+ \chi(t)\ \Phi_i(t) =\ H\delta_{i1},\label{phidot}
\end{eqnarray}
\begin{eqnarray}
\chi(t)= - \lambda v^2 +\lambda \Phi_i^2(t)
+ N\lambda\int\ddk ~|f_k(t)|^2(1+2\ n_k), \label{chinew}
\end{eqnarray}
\begin{eqnarray}
\ddot f_k +
\bigl( k^2 + \chi(t) \bigr) f_k .\label{fnew}
\end{eqnarray}

We also have the vacuum relations:
\begin{eqnarray}
m_{\pi}^2 = - \lambda v^2 +\lambda f_{\pi}^2
+ N\lambda\int\ddk ~ { 1 \over 2 \sqrt{k^2+m_{\pi}^2}} \label{mvac}
\end{eqnarray}

Which yields:
\begin{eqnarray}
\chi(t) &=&m_{\pi}^2+ \lambda \Phi_i^2(t) -\lambda f_{\pi}^2\nonumber \\ &&+
N\lambda\int\ddk ~[ |f_k(t)|^2(1+2\ n_k) - { 1 \over 2
\sqrt{k^2+m_{\pi}^2}}].\label{chivac}  \end{eqnarray}
These equations govern the growth of instabilities.
We see that there are unstable modes $f_k$ whenever $k^2 +\chi$ is negative so
that
only the long wave length modes can go unstable. If we ignore the
quantum fluctuation contribution,  $\chi$ can go negative
 only when
\begin{eqnarray}
  m_{\pi}^2+ \lambda \Phi_i^2(t) -\lambda f_{\pi}^2   < 0,\label{ineq}
\end{eqnarray}
so that the most unstable case has $\Phi =0$.One also has that the  bare
coupling
must  satisfy
\begin{eqnarray}
\lambda  > { m_{\pi}^2 \over  f_{\pi}^2} \label{ineq2}
\end{eqnarray}
for there to be any instabilities. Once instabilities grow, then
they cause exponential growth of the modes in the mode sum  which contributes
a positive quantity to the equation for $\chi$. Thus the stronger the coupling
the quicker $\chi$ returns to a positive value. In determining the parameters
of
our effective field theory we found that we are in the strong coupling regime
($\lambda_r >>1$  Thus we have the situation where any instabilities are
quickly
suppressed. This is not true at weak coupling such as found in the early
universe problems. In our approximation, we found that because of the
triviality
constraint
 the renormalized coupling constant is constrained to be
less than or equal to about 10 is we use $\Lambda= 1 GeV$. We merely comment
here
that if we were able to consider higher values of $\lambda_r$ then the
relaxation
of instabilities would be even faster and this would only dampen the
possibility of producing disoriented chiral condensates.

\section{Initial Conditions, Correlations and Domains}

\subsection{Reasonable initial conditions}
\medskip

How can we study
the possibility of forming large domains in which the pion field is
disoriented?
As we described above, the picture we have in mind is the
following: After the collision, a ``Baked Alaska'' type configuration is
formed.
When viewed in the natural ``boost invariant'' coordinate system this
configuration is described as a state characterized by homogeneous mean values
$\Phi_i(\tau)$
and by quantum and thermal fluctuations. The quantum state after the collision
is
clearly unknown and we will be forced to assume reasonable initial values
for the parameters consistent with being near local thermal equilibrium
in a disordered phase at a  time $\tau_0$ following the collision.  Ideally
we would like to be able to study the growth of inhomogeneous instabilities in
real space and time. However to simplify our calculation we are assuming that
the system evolves such that all expectation values just depend on the proper
time. This rules out a detailed study of domain growth and we have to
obtain information about the growth of domains indirectly from the
quantum correlation functions. These are parametrized by the proper time
evolving interpolating phase space number and pair densities.Thus, we will not
attempt to
describe here the process of formation of a ``peculiar'' quantum state
(which is
an interesting but very difficult issue). We will rather study the evolution of
different reasonable initial states constrained to have expectation
values which depend only on the proper time $\tau$, focusing on the possibility
of the development of instabilities and on the potential growth of long range
correlations. The question is how to choose the initial conditions defining a
``peculiar'', but  not entirely unrealistic initial state? Without any rigorous
justification, we  will assume an initial state that is a {\it disoriented (or
displaced)  thermal state}.
Let us describe it in more detail.

Let us forget for the moment about the expansion and consider a much simpler
situation:
thermal equilibrium at temperature $T$. This is characterized by a value of
$\chi=\chi_T$ and by mean fields $\vec\pi=\vec\pi_T=0$,
$\sigma=\sigma_T=H/\chi_T$. The value
of $\chi_T$, which is obtained by solving the gap equation given in the
appendix,  is positive and
fixes the ``effective mass'' of the quasiparticles (at very high
temperatures we have $\chi_T\propto\lambda_r T^2$ and $\sigma_T\propto H/T^2$,
i.e. the explicitly broken symmetry is restored at
high temperatures).

In choosing an initial state to model the situation after the collision
it is reasonable to assume the conditions of local thermal equilibrium above
the critical temperature so that we start in the disordered phase where the
chiral
symmetry is unbroken. In
particular, we can simply take the initial values $\sigma(\tau_0)=\sigma_T$,
$\vec\pi(\tau_0)=\vec\pi_T$ and $\chi(\tau_0)=\chi_T$ and turn on the
expansion at $\tau_0$ (the results of this simulation will be explained in
detail later). One should also consider the
possibility of exciting the initial state with
some extra kinetic energy. For this we should consider
an initial distribution of velocities $\dot\sigma(\tau_0)$
that kick
the initial mean values making it change in time
(it is worth noticing that the expansion destroys the initial
equilibrium and the fields
start to change even without any kick).
Other initial conditions we will consider are those in which the
initial state is a ``disoriented'' equilibrium where
$\vec\pi^2(\tau_0)+\sigma^2(\tau_0)=\sigma^2_T+\vec\pi^2_T$ but the
state is pointing in the wrong direction (say, along $\pi^0$). In fact,
the formation
of such disoriented
state is not at all unlikely: simple estimates show that the energy
required to ``tilt'' the initial state orienting it along the $\pi^0$ direction
(instead
of the $\sigma$ direction) is only $10\ MeV/fm^3$ \cite{BjorkenKT}.
Another possibility is to change the initial value of the
magnitude of the $O(4)$ vector making it different from its high temperature
equilibrium one. For all the above cases, the initial value
of $\chi$ will always be
determined by the solution of the initial gap equation (the same for its
initial
derivative) evaluated for an equilibrium configuration at a temperature $T_0$ .

However, we should point out that the initial value of $\chi$ will always
be restricted to be positive. As we said,
$\chi(\tau)$ is the effective mass of the quasiparticles. Therefore,
taking a negative initial value for $\chi$ implies that we are
``turning the effective potential upside down.'' In our view, this cannot be
the
consequence of forming a ``peculiar'' initial state but must rather be the
consequence of the cooling mechanism, which is entirely produced by the
expansion.
In fact, starting with a negative $\chi_0$ is what is done
when studying this problem by using the quench approximation: one starts with
a hot initial state and let it evolve in
the low temperature effective potential. It should be clear
by now that this is drastically different from our approach. We will study the
self-consistent evolution of $\chi$ starting from a ``hot'' initial value
and follow its evolution.

\subsection{Instabilities and correlations}

During the nonequilibrium evolution it is possible that, for certain time
intervals,
the value of $\chi(\tau)$ becomes negative (we will discuss some
examples below). When this happens there is an instability in the system.
As clear from (\ref{finaleqmode}), if $\chi\le 0$ there are long wavelength
modes that
become unstable: their amplitudes start growing exponentially
(the factor $1/4\tau^2$ very soon becomes negligible).
The existence of such unstable modes is the crucial ingredient needed
for the development of structure through
the mechanism of spinodal decomposition. Let us briefly describe now how we
will
study this issue here. The intuitive picture is clear: the system cools as it
expands
evolving
(in a fully nonequilibrium way) towards a stable low temperature state. In our
case,
such state (the vacuum) is characterized by the values $\vec\pi =0$,
$\sigma=f_\pi$ and
$\chi=m_\pi^2$. The existence of an instability means that the homogeneous
configuration is not energetically preferred and that any small inhomogeneity
seed
will tend to grow. As the $O(4)$ symmetry is explicitly broken, the growth of
structure is only transient since in the long run the stable state is again the
homogeneous vacuum. The question one should try to answer is if during the
nonequilibrium stage the instability is strong enough to form large domains in
which the field is correlated and disoriented. If such domains do form, the
correlations in the emitted pions can be detected. We should then
examine the evolution of
the correlation length (in the rapidity space).

A natural question that arises is how can one study the existence of
long range correlations within our scheme, which is
basically a mean field theory analysis. The textbook answer to this is that
mean field theory can still be used to provide useful information about the
typical scale of the correlations despite of the fact that the mean values are
taken as homogeneous \cite{Goldenfeld}. In fact, this analysis enables us to
compute the two point correlation functions which determine the behavior of
the system when perturbed away from its homogeneous configuration. Thus, the
correlation length obtained from the two point functions will characterize
the growth of structure, at least in the linear regime.
This point of view was used in \cite{BoyaPRD,BoyaPRE}).  Our feeling however
is that it is not so easy to interpret directly the two point correlation
function.  In the first order adiabatic basis we have:

\begin{eqnarray}
G^0(x,y;\tau)  = {1\over \tau} \int \ddk e^{ik(x-y)}
{1 \over 2 \omega_k(\tau)} \  (1+2\ n_\kk (\tau) + 2 Re [g_k(\tau)
e^{-i2y_k(\tau)}]) \label{adiadi}
\end{eqnarray}

We see that there are phases in the fourier transform $G(k,\tau)$ between
the number density $n_\kk (\tau)$  and the pair density $g_k(\tau)$ which can
make the interpretation difficult. These interpolating operators only make
physical sense when $<\pi>=0$ and for the stable modes. So at best we can look
at
the two quantities $n_k$ and $g_k$ and study their momentum dependence. If we
have a particular model for these quantities which have proper time dependent
masses as well as temperatures then  can one extract correlation
lengths (or  mass scales) in a model dependent fashion. For example if the
system expanded in local thermal equilibrium in a comoving frame we would
have:
\begin{eqnarray}
n_k(\tau) = {1 \over e^{\beta(\tau)  E _k(\tau)} -1},\label{loceq}
\end{eqnarray}
where $ \beta(\tau) = 1/T(\tau) $ and $E_k(\tau) = \sqrt{k^2+\chi(\tau)}$.

The effective temperature could be calculated by first determining the
hydrodynamical quantities $\epsilon(\tau)$ and $p(\tau)$ from the diagonal
entries of the expectation value of the energy momentum tensor in the comoving
frame as discussed in \cite{hydro}. Then one determines the temperature (and
entropy)
from the relations : $ \epsilon+ p = Ts$ ; $d \epsilon = T ds$.

We notice that this parametrization fails exactly when the pion mass goes
negative which is the case we want to study.  Also from our previous study
of the production of pairs from strong electric fields in a related
$1/N$ expansion \cite{hydro}, we know that the effective temperature in lowest
order
is not monotonically decreasing but is oscillating about a decreasing
function with the plasma oscillation frequency.  Thus until we add
scattering (which occurs at next order in the $1/N$ expansion), the temperature
parameter does not have this monotonicity property. At late times one expects
after a period of entropy production that $s \tau = constant$ and
$T(\tau) = T_0  ({\tau_0 \over \tau})^{1/3}$ as in the thermal equilibrium
case.
This is obviously one possible parametrization of the data in terms of two
$\tau$ dependent inverse length scales, the temperature and the mass of the
pion. As we shall discover, this parametrization of the number density does
not agree with our nonequilibrium evolution so that there are at least 3 proper
time evolving
length scales in this problem--the inverse mass of the pion, the inverse of
the effective temperature and possible length scales describing domain
growth.
 In what follows we will
 present our results for these two interpolating densities to see their
general  proper time development without making specific models in terms
of various length scales except for the quasithermal model which does not
reproduce the results.

The computation of the pion two point function in general is straight forward.
The two point function is obtained by inverting the inverse propagator
matrix as discussed in the appendix. In general we get for the inverse
of the pion two point function:
\begin{eqnarray}
G^{-1}_{\pi \pi} (x,y)_{ij} = -\delta_{ij} [\Box + \chi(x)]\delta(x-y)
-\pi_i(x) G_{0 \chi \chi} (x-y) \pi_j(y)
\end{eqnarray}
We see that only  when $\pi_i(x)$ = 0 can one obtain the Green's function
directly in terms of the modes used to calculate the quantum field $\phi$. Also
only in that case do we get a direct interpretation of the Fourier transform of
the Green's function in terms of the (proper) time dependent interpolating
phase
space number and pair densities described above.
As discussed earlier it is simple to extract $n$ and $g$ from the solution
$f$ of the mode equation.

Although the equal proper time correlation function depends only
on the time dependent number and pair densities, the full non-equal time
correlation function has more phase information and is given by
(when $<\pi> =0$):

\begin{eqnarray}
G^0(x_{\perp}-x^{\prime} _{\perp};\eta-\eta^{\prime};\tau, \tau^{\prime})  =
{1 \over \sqrt{\tau }} {1 \over
\sqrt{\tau^{\prime}}} \int \ddk e^{-ik(x-x')}
\end{eqnarray}
\begin{eqnarray}
 \{(1+2\ n_\kk)Re
[f^{\ast}_k (\tau) f_k(\tau')] + 2
Re [g_k f_k (\tau) f_k(\tau')]\} \label{fullg}
\end{eqnarray}
where here $n,g$ refer to their value at $\tau_0$.

The only thing left to discuss is the choice of initial proper time $\tau$.
One can estimate this quantity in a hydrodynamical model with Landau's
choice of initial conditions (namely energy density of a Lorentz contracted
pancake being given at an time zero).
If one assumes that
the the collision produces a quark-gluon plasma in equilibrium, one can
run the hydrodynamical code for initial energy densities appropriate to
the collision of Lorentz contracted disks and determine the proper
time when the critical temperature is around $200 MeV$ or slightly
above the chiral phase transition. We want to start in the quark - gluon
plasma phase above the chiral phase transition which places constraints on the
initial energy density
present at $\tau_0$. We then assume that slightly above this transition
it is reasonable to model the chiral transition with the effective Lagrangian
of the sigma model. Not knowing exactly what this proper time is, we
will here consider reasonable initial proper times ($1 fm/c < \tau_0 <4 fm/c$
and
study the effect of the initial proper time $\tau$ on the production of
instabilities.
  Taking
larger proper times as the  starting point for our calculation
reduces even further the possibility of instability growth so that our results
present an upper limit on the growth of domain sizes in this model.

\section{Results}

We have performed numerical simulations on the connection machine CM-5 using
a grid that has $10000 \tau$ modes. (We start at $\tau_0 = 1$). The grid
is 100 modes wide in the transverse direction and $ 100 {\ tau \over \tau_0}$
modes in the
$\eta$ direction.  That is we choose:
$$
dk_{\perp} = { \Lambda \over 100}, \hspace{.2in} {dk_{\eta} \over 1fm} =
{\Lambda \over 100}. $$
The first issue we will examine here is the existence of instabilities. In
previous works the presence of unstable modes was assumed by imposing the
quench
approximation. In  our numerical studies we find that, due to the strong
coupling, that a wide class of initial conditions consistent with reasonable
fluctuations found in a thermal distribution no instabilities develop.  In
our simulations the one thing we did not change was the value of the
composite field $\chi$ which was fixed to be the solution of the gap equation
in the initial thermal state. We also maintained the constraint that
the initial values of the expectation values of the field, namely $\pi(\tau_0)$
and $\sigma(\tau_0)$ satisfied the relationship:
$$\vec\pi^2(\tau_0)+\sigma^2(\tau_0)=\sigma^2_T
$$
 where $\sigma_T$ is the
equilibrium value of $\Phi$ at the initial temperature $T$. This last
constraint
is reasonable since it doe not cause much energy to make a rotation in the
direction of $\pi_i$.
Thus we chose different initial $\sigma$ and $\pi_i$ consistent with the above
constraints and then probed the effect of different initial values for
$\dot{\sigma}$ and $\dot{\pi}_i$.  Since the results for $\dot{\pi}_i \neq 0$
were similar to those when $\dot{\sigma} \neq 0$ we mainly present here
the results for the latter case.
We have also surveyed other possibilities that violate the above constraints
such
as choosing initial $\sigma=0$. In that case we found no instabilities
even with $ \dot{\sigma} \neq 0$.

First let us consider  the case when the initial value $\sigma(\tau_0)$ is
the thermal equilibrium one corresponding to a temperature of $200 MeV$. We
varied the value of the initial proper time derivative of the sigma field
expectation value and found that there is a narrow range of initial values that
lead to the growth of instabilities. Namely
$$  .25 <  \vert \dot{\sigma} \vert   < 1.3
$$
 Surprisingly when $ \vert \dot{\sigma} \vert > 1.3 $ instabilities no longer
occur.

Figures 1-2 summarize the results of the numerical simulation for the
evolution of the system (\ref{finaleqphi})--(\ref{finaleqmode}).
We display the auxiliary field $\chi$ in units of $fm^{-2}$  , the classical
fields $\Phi$ in units of $fm^{-1}$ and the proper time
in units of $fm^{-1} $   ($ 1fm^{-1} = 197
MeV$).
In Fig.~1 and Fig.~2 the proper time evolutions of the auxiliary field $\chi$
field are presented for two different values of $f_{\pi}$,
where the initial conditions were fixed at $\tau=1$. The initial conditions in
all the cases are with the equilibrium value of the $\chi$ field at the
corresponding initial temperature, where as the various initial conditions of
the
classical fields $\Phi$ are chosen to satisfy
$\vec\pi^2(\tau_0)+\sigma^2(\tau_0)=\sigma^2_T$ where $\sigma^2_T$ is the
equilibrium value of $\Phi$ at the initial temperature $T$. In the case of
$f_{\pi}= 92.5 MeV$ we see that the instability lasts for less than $3 fm$. As
discussed earlier, the regime of exponential growth of the unstable modes
occurs
whenever $\chi < 0$. We notice that when we rotate the expectation value from
the $\sigma$ to the $\pi^1$ direction initially then the instability regime has
the same typical ``survival'' time.  We also notice that if we choose the time
derivative  to
be zero  and start from an equilibrium configuration, then the expansion alone
is insufficient to generate instabilities.
We find that in order to generate instabilities we require
fluctuations in the classical kinetic terms such as $\dot{\sigma}$ or
$\dot{\pi}_i$ and as discussed earlier  these initial conditions must be in
a very narrow range to produce instabilities. Thus the rapid quench conditions
assumed by other authors comprises only a small region of the phase space
of initial fluctuations expected in an initial thermal distribution.
 As
expected, at late proper times, the auxiliary field approaches its equilibrium
value of $m_{\pi}^2$.   For the case when $f_{\pi}=125 MeV$, a value favored by
fitting the low energy scattering, we  see a very  similar behavior, showing
that our main results concerning the small range of initial conditions that
lead to instabilities are not affected by a $30\%$ change in the value of one
of our parameters.

\begin {figure*}
 \centering{\
  \epsfysize=10.0cm
  \epsfxsize=12.0cm
}
\caption{Proper time evolution of the $\chi$ field
for four different initial conditions with $f_{\pi}=92.5 MeV$.}
\end {figure*}
\begin {figure*}
\centering{\
  \epsfysize=10.0cm
  \epsfxsize=12.0cm
}
\caption{Same as Fig.~1, but for $f_{\pi}=125 MeV$.}
\end {figure*}
A crude estimate for the size of a disoriented chiral condensate is to
multiply a typical survival time length for the instabilities  by the speed of
light. If this
estimate is valid then the size of these regions are of the
order of a few fermi. Of course one needs to study the
growth of inhomogeneous instabilities to make any definite statements
about this size.  We were hoping that the correlation functions we have
calculated
could be interpreted easily in terms of a length parameter associated
with the size of these regions. However as we will see below
such a naive hope is not to be satisfied.
In Fig. 3 we plot the proper time evolution of the classical fields
$\sigma$ and $\pi$ for the same two choices for $f_{\pi}$ as in Fig.~1 and
Fig.~2.
We see that the sigma field asymptotes to its vacuum value $f_{\pi}$ and the
$\pi$ field gradually converges to its equilibrium value of zero.
\begin {figure*}
\centering{\
  \epsfysize=10.0cm
  \epsfxsize=12.0cm
}
\caption{Proper time evolution of the $\sigma$ and $\pi$ fields
for the following initial conditions:
Solid line is for $\sigma(1)=\sigma_T$, $\pi^i(1)=0$ and $\dot{\sigma}(1)=1$.
Dotted line is for $\sigma(1)=\sigma_T$, $\pi^i(1)=0$ and $\dot{\sigma}(1)=0$.
Dashed- dotted line is for $\sigma(1)=\sigma_T$, $\pi^i(1)=0$ and
$\dot{\sigma}(1)=-1$.  Dashed line is for $\sigma(1)=0$, $\pi^i(1)=\sigma_T$
and
$\dot{\sigma}(1)=-1$.  At $T=200 MeV$, $\sigma_T= 0.3 fm^{-1}$ for
$f_{\pi}=92.5 MeV$  and $\sigma_T= 0.5 fm^{-1}$ for
$f_{\pi}=92.5 MeV$ }
\end {figure*}
In Fig. 4 we show the effect of starting the initial value problem
at later times, namely $\tau_0=2.5,4 fm$ and compare them to the case
previously
studied with $\tau_0= 1 fm$ which had a modest region of instability growth.
We see that as we increase $\tau_0$ we decrease the possibility of instability
growth and by these late times it is not possible to produce instabilities even
with kinetic energy fluctuations.
\begin {figure*}
\centering{\
  \epsfysize=10.0cm
  \epsfxsize=12.0cm
}
\caption{Proper time evolution of the $\chi$ field
for three different initial proper times $\tau_0=1,2.5,4$ with $f_{\pi}=92.5
MeV$, $\sigma(\tau_{in})=\sigma_T$, $\pi^i(\tau_{in})=0$ and
$\dot{\sigma}(\tau_{in})=-1$.} \end {figure*} In Fig. 5 we study the effect of
the initial temperature on our time evolution problem. For $f_\pi = 92.5$ the
critical temperature is $160 MeV$ in the absence of explicit symmetry
breaking. We see that in the vicinity of the critical
temperature the effects of varying the initial temperature is minor.
\begin {figure*}
\centering{\
  \epsfysize=10.0cm
  \epsfxsize=12.0cm
}
\caption{Proper time evolution of the $\chi$ field
for three different initial thermal distributions with $T=200,164,150 MeV$ for
the initial conditions $\sigma(\tau_{in})=\sigma_T$, $\pi^i(\tau_{in})=0$ and
$\dot{\sigma}(\tau_{in})=-1$.} \end {figure*}
In Fig. 6 we study the proper time evolution of the effective number density
for various initial conditions. In thermal equilibrium one expects for
an isentropic expansion in boost invariant coordinates that
$s \tau= constant$. Since the number density is proportional to the entropy
density one expects that once particle production stops that
$n(\tau) \tau \rightarrow constant$.  We see this trend in this figure, showing
that we are reaching
the out regime as the system expands. The breaks in some of these
curves at early values of $\tau$ are a result of the fact that the
interpolating number density cannot be defined when $ \chi$ is negative.
\begin {figure*}
\centering{\
  \epsfysize=10.0cm
  \epsfxsize=12.0cm
}
\caption{Proper time evolution of the produced particle density $n\tau\equiv
dN/d\eta dx_{\perp}$
for the same evolution shown in Fig.~1}
\end {figure*}
Next we are interested in knowing how our results differ from the case where
the system evolves in local thermal equilibrium
which is described by two correlation lengths, the inverse of the effective
pion
mass  associated with $\chi$ , and the inverse of the proper time evolving
effective temperature $ T(\tau) = T_0 ({\tau_0 \over \tau})^{1/3}$ discussed
earlier. We see from Fig.~7 that
in the case that $\sigma(1)=\sigma_T$, $\pi^i(1)=0$ and $\dot{\sigma}(1)=-1$,
where maximum instability exists, complex structures are formed as contrasted
to
the local thermal equilibrium evolution. The interpolating phase space
distribution $n(k_{\eta},{\bf{k}}_{\perp},\tau)$ obtained numerically, clearly
exhibits a larger correlation length in the transverse direction than the
equilibrium one and has correlation in rapidity of the order of 1-2 units of
rapidity. We notice that in both directions there is structure which does not
lend itself to a simple interpretation. On the other hand the  local thermal
equilibrium evolution is quite regular apart from the normalization of the
distributions that are changing with time due to oscillation in the quantity
$\chi(\tau)$ which is damped to its equilibrium value once the system expands
sufficiently. Other authors have suggested that it is possible to extract the
sized of the domains of DCC's from the coordinate space correlation function
$G(x,y,\tau)$  by considering the width of the distribution as a measure of the
size of the domain.
However, it is clear from our detailed numerical results that there are several
length scales affecting the momentum space distribution apart from the
effective
pion mass and temperature.  The spatial Green's  function
depends on not only $n_k$ and $g_k$ but also the phases $y_k$ and would be even
harder to interpret than $n$ and $g$ separately.
\begin {figure*}
\centering{\
  \epsfysize=10.0cm
  \epsfxsize=12.0cm
}
\caption{Slices of $k_{\eta}=0$ and $p\equiv\vert{\bf{k}}_{\perp}\vert=0$ of
the proper time evolution of the interpolating  phase space  particle number
density $n(k_{\eta},{\bf{k}}_{\perp},\tau)$   for $\sigma(1)=\sigma_T$,
$\pi^i(1)=0$ and $\dot{\sigma}(1)=-1$ compared with the corresponding local
thermal equilibrium densities $n_T (k_{\eta},{\bf{k}}_{\perp},\tau)$.}
\end {figure*}
In Fig. 8 we look at the time evolution of the interpolating number
operator for the case when we did not have any fluctuations in
the kinetic energy. Here we are still far from equilibrium and we see that
unlike the equilibrium case the correlation length in rapidity space is
not decreasing with proper time. In this case where there are no
instabilities we do not see complicated structures. The transverse
distribution is similar to the equilibrium case.
\begin {figure*}
\centering{\
  \epsfysize=10.0cm
  \epsfxsize=12.0cm
}
\caption{As in Fig.~7, but
for $\sigma(1)=\sigma_T$, $\pi^i(1)=0$ and $\dot{\sigma}(1)=0$.}
\end {figure*}
The pair density function $g(k_{\eta},{\bf{k}}_{\perp},\tau)$ is even more
elusive to parametrize than
the single particle distribution function. In the case where there are no
instabilities ($\dot{\sigma}(1)=0$) we see in Fig.~9 that
although the transverse distribution is relatively simple, the
distribution in the $\eta$ direction (whose fourier transform gives
the rapidity distribution) has many length scales.  When we have
instabilities ($\dot{\sigma}(1)=-1$) then both distributions are
complicated and possess several length scales as seen in the lower part of
Fig.~9.
\begin {figure*}
\centering{\
  \epsfysize=10.0cm
  \epsfxsize=12.0cm
}
\caption{Slices of $k_{\eta}=0$ and $p\equiv\vert{\bf{k}}_{\perp}\vert=0$ of
the proper time evolution of the real part of the phase space pair density
$g(k_{\eta},{\bf{k}}_{\perp},\tau)$   for $\sigma(1)=\sigma_T$, $\pi^i(1)=0$
and
$\dot{\sigma}(1)=0,-1$.} \end {figure*}

\section{Conclusions}

In this paper we have performed numerical simulations in the regime
of the chiral phase transition in the linear sigma model for a wide
variety of initial conditions starting above the critical temperature for
an expanding plasma of pions and sigmas. Assuming that this model gives
a reasonable description in this temperature range, we found initial conditions
where instabilities grew in the scenario required for the formation of
disoriented chiral condensates.  The constraints on our model which are imposed
in order to approximate the low energy physics of pion interactions however
require a large renormalized coupling constant. This had the effect of
rapidly damping the instabilities and we found no evidence in this model
for large domains of disoriented chiral condensates. We did, however see rather
large departures in the phase space number density from one which would result
from an evolution in local thermal equilibrium. These departures show a
narrowing of the momentum space distribution which could be interpreted as
a larger spatial correlation length. However we found no simple method of
extracting correlation lengths from our results. In our simulations we assumed
a
reasonable mechanism for cooling-- namely the expansion of the plasma following
its production in a collision. However we did ignore scattering, which occurs
at next order in the $1/N$ expansion.We also did not allow for general
inhomogeneous fluctuations which would allow us to actually study the growth
of disoriented domains. This would require a study of the growth
of inhomogeneous perturbations about our homogeneous background and would
involve a much more difficult numerical computation. We hope to
carry out such a computation in the future. It is also possible that
another model, such as the non-linear sigma model might better describe the
dynamics in this regime and might lead to different conclusions from those
found here.  In the future we also hope to include  scattering effects which
will introduce another time scale, namely the equilibration time scale
into the problem, as well as
study the nonlinear sigma model to see if the results differ significantly from
those found here.

\section{ Appendix I- $\sigma$ model in the large $N$ approximation}
The $O(4)$ $\sigma$ model is described by the Lagrangian
\begin{eqnarray}
L= \{{1\over 2} \partial\Phi \cdot \partial\Phi - {1\over 4}
\lambda (\Phi \cdot \Phi - v^2)^2 + H\sigma\}.\label{action2}
\end{eqnarray}
where the mesons are organized in an $O(4)$ vector
$\Phi=(\vec\pi,\sigma)$ . The counting of the large N expansion is made
explicit \cite{1/n} by introducing a composite field $\chi$ defined by
$\chi = \lambda (\Phi \cdot \Phi-v^2)$. It is easy to show by appropriate
rescalings
that the large N expansion is obtained by integrating out the $\Phi$ field
and then performing a steepest descent calculation on the remaining $\chi$
path integral. That is the large N expansion is a loop expansion in the
composite field $\chi$ propagator \cite{higher}. Thus we add to the above
Lagrangian the constraint equation:
\begin{eqnarray}
{[\chi - \lambda (\Phi \cdot \Phi-v^2)]^2 \over 4 \lambda},\label{const}
\end{eqnarray}
which exactly cancels the quadratic term. We thus obtain the alternate
Lagrangian:
\begin{eqnarray}
L_2 = -{ 1 \over 2} \phi_i (\Box + \chi) \phi_i + {\chi^2 \over 4 \lambda} +
{1 \over 2} \chi v^2 + H \sigma. \label{alterl}
\end{eqnarray}

The generating functional of the Green's functions  is given by the Path
Integral:
\begin{eqnarray}
Z[J ,S] = \int D \chi ~ D \Phi ~ {\rm exp}~\{ i~\int d^4x [L_2 + J \cdot \Phi+
S
\chi]\}] \equiv {\rm exp}[i W(J,\chi)] .\label{generate}
\end{eqnarray}
We can now perform the Gaussian path integral over the $\Phi$ field. Evaluating
the remaining $\chi$ integral at the stationary phase point of the resulting
effective action, and then Legendre transforming:
\begin{eqnarray}
\Gamma[\Phi,\chi] = W[J,S] - \int d^4x [J(x) \cdot \Phi(x) + S(x) \chi
(x)],\label{effacc}
\end{eqnarray}
we obtain the lowest order (in large $N$) result:
\begin{eqnarray}
\Gamma[\Phi,\chi] = \int d^4x[ L_2(\Phi,\chi,H) + { i \over 2}N {\rm tr~ln}
G_0^{-1}],\label{effac2}
\end{eqnarray}
where
\begin{eqnarray}
G_0^{-1}(x,y) = i[\Box + \chi(x)] ~\delta^4(x-y). \label{ginvapp}
\end{eqnarray}
For the $O(4)$ sigma model $N=4$.
{}From this effective action we immediately get the equations of motion for
the fields $\Phi$ and the equation of constraint (gap equation) for the
composite field $\chi$ :
\begin{eqnarray}
[\Box + \chi(x)] \pi_i = 0 ~~~~ [\Box + \chi(x)]\sigma = H,
\end{eqnarray}

\begin{eqnarray}
 \chi= - \lambda v^2 + \lambda (\sigma^2 + \pi \cdot \pi) + \lambda
N  G_0 (x,x). \label{chiapp}
\end{eqnarray}

In the static case when we are considering symmetry breaking, we
obtain lowest energy state when:

\begin{eqnarray}
 \chi\sigma = H ; ~~~  \chi= - \lambda v^2 + \lambda \sigma^2 +
\lambda N  G_0 \label{vacapp}.
\end{eqnarray}

By considering the fields $ ( \Phi,\chi )$ to be a 5 dimensional vector $\Psi$,
and the matrix inverse propagator to be:
\begin{eqnarray}
\hat{G}^{-1} =  {\delta^2  \Gamma \over \delta \Psi \delta \Psi}
,\label{invapp}
\end{eqnarray}
we obtain by inverting this matrix the relevant $\pi, \sigma$, and
$\chi$ propagators.  Performing this inversion we obtain for the vacuum
Feynman
$\chi$ inverse propagator:
\begin{eqnarray}
\hat{G}^{-1}_{\chi \chi}(p^2)= G^{-1}_{0 ~\chi \chi}(p^2)+ {\sigma^2 \over
\chi-p^2},\label{chiinvapp}
\end{eqnarray}
where
\begin{eqnarray}
{\hat{G}^{-1}}_{0 ~ \chi \chi}(p^2) = { 1 \over 2 \lambda} + {N \over 2}
\Pi(p^2)
\end{eqnarray}
is the inverse propagator in the absence of symmetry breaking and the
polarization $\Pi = i G_{0}^2$ is given by
\begin{eqnarray}
\Pi(p^2) = -i \int ~[d^4q] ~
(\chi-q^2)^{-1} (\chi-(p+q)^2)^{-1}. \label{polapp}
\end{eqnarray}

In our simulations we have a three dimensional cutoff. If we integrate
over $q^{0}$ and perform the remaining three dimensional integral with
a cutoff $\Lambda$ we obtain explicitly:
\begin{eqnarray}
\Pi(p^2)= {1 \over 8 \pi^2}\left[{\rm \ln} ({1 \over x} + \sqrt{1+{1\over
x^2}})
+{1 \over 2} \sqrt{1- {4m^2 \over p^2}} {\rm \ln} \left({  \sqrt{(1-
{4m^2 \over p^2})(1+x^2)}-1 \over  1+ \sqrt{(1-{4m^2 \over p^2})(1+x^2)}}
\right) \right]\label{pol2app} \nonumber \\
\end{eqnarray}
where
$x=m/\Lambda$ and $m$ is the pion mass: $\chi=m^2$.

For the pion inverse propagator we obtain in the vacuum sector:
\begin{eqnarray}
{\hat{G}^{-1}} _{\pi \pi}(p^2)_{ij} = \delta_{ij} (p^2-\chi)=\delta_{ij}
(p^2-m^2). \label{pipropapp}
\end{eqnarray}
Thus we see that $\chi= m^2$ is the pion mass squared.
In our initial value problem one can have $\pi_i \neq 0$. For that case
(or when there are external isospin sources present), the pion inverse
propagator
becomes for constant external sources:
\begin{eqnarray}
{\hat{G}^{-1}} _{\pi \pi}(p^2)_{ij} = \delta_{ij} (p^2-\chi)
- \pi_i  \hat{G}_{0 ~ \chi \chi}(p^2) \pi_j. \label{constapp}
\end{eqnarray}

For the $\sigma$ inverse propagator we obtain:
\begin{eqnarray}
{\hat{G}^{-1}} _{\sigma \sigma}(p^2)= p^2-m^2 - \sigma^2 \hat{G}_{0 ~ \chi
\chi}(p^2).\label{sigmapropapp}
\end{eqnarray}
Thus the $\sigma$ mass $m_{\sigma}^2$ is determined by the relationship:

\begin{eqnarray}
m_{\sigma}^2=m^2 +\sigma^2 Re[\hat{G}_{0 ~\chi \chi}(m_{\sigma}^2)].
\label{simaapp}
\end{eqnarray}
 The axial current in this model is given by :
\begin{eqnarray}
 A_{\mu}^i (x) = [ \pi^i \partial_{\mu} \sigma -
\sigma \partial_{\mu} \pi^i] . \label{pcac}
\end{eqnarray}
The PCAC condition  is given by:
\begin{eqnarray}
\partial_{\mu} A_{\mu}^i (x) = H  \pi^i(x). \label{pcac2}
\end{eqnarray}
Therefore one has
\begin{eqnarray}
   H= f_{\pi} m^2. \label{happ}
\end{eqnarray}
Since we also have that the vacuum is defined by $ \chi \sigma= m^2 \sigma=H$,
we immediately find that $\sigma=f_{\pi}$.  We can rewrite
the gap equation in terms of physical quantities as:
\begin{eqnarray}
m^2 = - \lambda v^2 + \lambda f_{\pi}^2 + \lambda N G_0 (x,x), \label{gappap}
\end{eqnarray}
where
\begin{eqnarray}
G_0(x,x) =\int_0^{\Lambda} \ddk {1\over 2\sqrt{k^2+m^2}}
\end{eqnarray}
This relationship allows us to determine the bare mass $-\lambda v^2 $
in terms  of the renormalized mass  m, the pion decay constant $f_{\pi}$ and
the
cutoff $\Lambda$

To determine $\lambda_r$
we must look at either the $\sigma$ mass which is not well determined by
experiment, or the low energy scattering data such as the s wave scattering I=0
scattering amplitude. Although at tree level
 the sigma mass is an arbitrary parameter, that is not true
for the quantum theory.  The reason for this is that we only want
to consider this model as an effective field theory with a cutoff, so
that the bare coupling constant is always positive. (Renormalized ordinary
perturbation theory requires that the bare coupling constant is negative as
we take away the cutoff). Once there is a cutoff, then the property of
the exact $O(N)$ sigma model, as determined by lattice or strong
coupling expansions is that the renormalized coupling is a monotonically
increasing function of the bare coupling reaching a finite maximum as the
bare coupling goes to infinity. In four dimensions, this maximum value
decreases logarithmically with the cutoff. This exact feature of the $O(N)$
sigma model is preserved in the large N approximation which is a virtue
of our approximation. However for reasonable cutoff (say 1 Gev)
this puts and upper bound on the $\sigma$ mass of around $3 m$ as we shall
see below.
  The low energy scattering data presents another problem. This data is
quite difficult to  fit in perturbation theory because perturbative
results violate partial wave unitarity.  Basdevant and Lee \cite{BasLee} were
able  to fit the s wave data by first calculating to one loop and then
enforcing unitarity by doing a Pade approximant.  Our $1/N$  approximation
is related to the Pade approximant since it sums the bubbles of the
perturbative result, but it does not automatically obey partial wave unitarity.
Nevertheless it is important to see what values of $\lambda_r$ give reasonable
s wave phase scattering amplitudes in our approximation.
  Let us therefore turn our attention to determining the low energy scattering
amplitude in the large N limit.
The pion-pion scattering invariant T-matrix is given by:
\begin{equation}
T= \delta_{ij} \delta_{kl} A(s) + \delta_{ik} \delta_{jl} A(t)
+ \delta_{il} \delta_{jk} A(u)
\end{equation}
where the isospin indices $i, j, k, l$ are coupled to the four
momenta $p_1, p_2, p_3, p_4$ such that $s=(p_1+p_2)^2$, $ t= (p_1-p_3)^2$
$u=(p_1-p_4)^2$. In leading order in large $N$ the amplitudes $A(s)$, $A(t)$
and
$A(u)$ are
exactly the $\chi$ propagator. Namely
\begin{equation}
A(s) = -\hat{G}_{\chi \chi}(p^2=s).
\end{equation}
The large N expansion preserves the current algebra so that
the usual low energy theorem is exact. That is, for small $s, m^2$ one
easily shows that
$$ A(s) \rightarrow {(s-m^2) \over f_{\pi}^2}.$$
This amplitude is independent of the coupling constant $\lambda$.
We thus need to go above threshold to determine the coupling constant.
The $I=0$ scattering amplitude is
\begin{eqnarray}
A^0=3A(s)+A(t)+A(u).
\end{eqnarray}
  The s-wave scattering amplitude is obtained by integrating the
the I=0 scattering amplitude over angles.
\begin{eqnarray}
f_{l=0} =e^{i \delta(s)}{\rm sin} ~\delta(s) = {1 \over 32 \pi} \sqrt{1-{4m^2
\over s}} \int_{-1} ^{1} dz ~ A^0, \label{swave}
\end{eqnarray}
where  $z={\rm cos} ~\theta$ and $\theta$ is the scattering angle in
the s channel center of mass system.

  It is useful to describe the theory in terms of the renormalized
coupling constant $\lambda_r$ which depends on $\lambda$ as well as the cutoff
$\Lambda$. The running renormalized coupling constant is determined by the
renormalization group invariant composite field propagator $G_{\chi~\chi}$.We
choose to define the renormalized coupling constant $\lambda_r$ as the running
coupling
constant  at $q^2=0$, for the unbroken mode of the theory. That is:
\begin{eqnarray}
\lambda_r ={\lambda \over  1+ N \lambda \Pi(q^2=0)}
\end{eqnarray}
Rewriting the propagator in terms of $\lambda_r$,one immediately gets a finite
expression for the running coupling constant.
\begin{eqnarray}
2 \lambda_r(q^2) =  \hat{G}_{\chi~\chi}(q^2) ={2\lambda_r \over
1+ \lambda_r N \Pi_r(q^2)-{2 \lambda_r f_{\pi}^2 \over
q^2-m^2}  } ~~, \label{running}
\end{eqnarray}
with
\begin{eqnarray}
\Pi_r(q^2)= \Pi(q^2) - \Pi(q^2=0). \label{pirenorm}
\end{eqnarray}
In Fig. 11    we  plot the real part of the s wave scattering
amplitude   for $\lambda_r=7.3 ,f_{\pi}=125 MeV$ and $\lambda_r=7.3 ,
f_{\pi}=92.5 MeV$.
By comparing these curves to the unitarized perturbation theory results of
Basdevant and Lee \cite{BasLee} we find reasonable agreement in the regime
$ 2< \sqrt{s/m} < 2.6$ for $f_{\pi}=125 MeV$ .This value of $f_{\pi}$
is obviously the preferred one if we want a closer agreement of our results
with those of Basdevant and Lee. (However since in our large N expansion we
have
only a modest value of N, namely $ N=4$
we might expect $25\%$ corrections at next order in $1/N$).  We
notice the existence of the sigma resonance at a mass of around 3m. These
values
of $f_{\pi}$ and $m_{\sigma}$ are in reasonable agreement with the fits of
Basdevant and Lee using their Pade analysis of ordinary perturbation theory.
 We  see that in our approximation there is a slight break down in
s-wave unitarity near the peak.(s-wave unitarity requires that the real
part of this
amplitude to be less than $1/2$). This breakdown in s-wave unitarity occurs in
the leading order in large-N  approximation for renormalized couplings which
are
larger than three.
\begin {figure*}
\centering{\
  \epsfysize=15.0cm
  \epsfxsize=12.0cm
}
\caption{The real part of the partial wave amplitude $f_{l=0}$ for $I=0$
computed from the linear $\sigma$-model in the lowest order of the $1/N$
expansion
for $f_{\pi}=92.5 MeV$ and for $f_{\pi}=125 MeV$. The value of $\lambda_r$ used
was $7.3$.}
\end {figure*}

One can approximately calculate the sigma mass from the zero of the
real part of the inverse propagator. In terms of the renormalized coupling
defined above we have:
\begin{eqnarray}
m_{\sigma}^2 = m^2 + Re[{ 2 \lambda_{r} f_{\pi}^2 \over 1 + \lambda_r N
\Pi_r(m_{\sigma}^2)}] . \label{massapprox}
\end{eqnarray}
This equation,which leads to  a similar value for the mass of the sigma as
found from the peak in the scattering,
 shows that there is an upper bound on the sigma mass which depends
on the chosen cutoff, since $\lambda_r$ decreases monotonically
with the cutoff.

The phase structure of the $\sigma$ model in this approximation
has been studied extensively by W. Bardeen and Moshe Moshe \cite{cutoff}.  For
zero temperature the expectation value  $ < T_{00}>$ in our initial density
matrix is the correct energy density functional to study. At finite
temperature it is the free energy density expectation value that is needed to
determine the phase structure.   Explicitly at zero temperature one has that
the
energy density functional is: \begin{eqnarray}
W(\Phi \cdot \Phi, \chi) = {N \over 64 \pi^2} \chi^2 {\rm ln} [{e^{1/2}
\Lambda^2 \over
\chi}] + {\lambda \over 4} [ \Phi \cdot \Phi - v^2 - {N \over 16 \pi^2} \chi
{\rm ln} ({e \Lambda^2 \over \chi})]^2. \label{finitetemp1}
\end{eqnarray}
Minimizing this equation with respect to $\chi$ gives the gap equation
\begin{eqnarray}
\chi = \lambda (\Phi \cdot \Phi-f_{\pi} ^2) - {N \lambda \chi \over 16 \pi^2}
{\rm ln}({e \Lambda^2 \over \chi}), \label{finitetemp2}
\end{eqnarray}
 which
then gives $\chi$ as a functional of $\Phi \cdot \Phi$. This then implicitly
 determines the energy functional as a functional of only $\Phi \cdot \Phi$.
Bardeen and Moshe Moshe point out that if one uses the endpoint solution
$\chi=0$ below the minimum of the potential, then one gets a real ``effective
potential".

 At finite temperature the phase structure of the cutoff theory is determined
instead by the free energy density functional:

\begin{eqnarray}
W(\Phi \cdot \Phi, \chi)&=&{N \over 64 \pi^2} \chi^2 {\rm ln} [{e^{1/2}
\Lambda^2 \over \chi}] - {N \over 48} T^4 \int_0 ^ {\chi /T^2} dy ~ y [{dF(y)
\over dy}] \nonumber \\
&&+ {\lambda \over 4} [ \Phi \cdot \Phi - v^2 + {N \over 16 \pi^2} \Lambda^2-
{N
\over 16 \pi^2} \chi  {\rm ln} ({e \Lambda^2 \over \chi}) +{N \over 24} T^2
F({\chi \over T^2})]^2  , \nonumber \\  \label{freeenergy}
\end{eqnarray}
where
\begin{eqnarray}
F(x) = {6 \over \pi^2} \int_0^{\infty} ~ {dy~y^2 \over \sqrt{y^2+x}}
[{\rm exp}( \sqrt{y^2+x} )-1]^{-1}. \label{fx}
\end{eqnarray}
The gap equation is now:
\begin{eqnarray}
\chi = \lambda (\Phi \cdot \Phi-f_{\pi} ^2) - {N \lambda \chi \over 16 \pi^2}
{\rm ln}({e \Lambda^2 \over \chi}) + {\lambda N \over 12} T^2 F({\chi \over
T^2}).\label{gapgap}
\end{eqnarray}

Using this free energy one finds that the critical temperature is determined
from:
\begin{eqnarray}
{N T_c^2 \over 12} = f_{\pi}^2. \label{critapp}
\end{eqnarray}

These relations are used to start our calculation above $T_c$.

\section{Acknowledgements}
 We would like to thank our colleagues at Los Alamos, especially Alex Kovner
for the many enlightening discussions we had, which were very valuable
and Salman Habib for stimulating criticisms and discussions. One of us
(F.C.) would like to thank Bjorken for encouraging us to do this
calculation. We also benefitted from several discussions with D. Boyanovsky and
Ian Kogan.

\end{document}